\documentclass[doublecol]{epl2} 
\usepackage{braket}
\begin{document}
\title{Evaporative cooling and self-thermalization in an open system of interacting fermions }

\author{Andrey R. Kolovsky\inst{1,2} \and  Dima L.Shepelyansky\inst{3}} 
\shortauthor{A.R.Kolovsky, D.L.Shepelyansky}
\shorttitle{Self-thermalization in open system of weakly interacting fermions}
\institute{
 \inst{1}  Kirensky Institute of Physics, 660036 Krasnoyarsk, Russia\\
 \inst{2} Siberian Federal University, 660041 Krasnoyarsk, Russia\\
 \inst{3} Laboratoire de Physique Th\'eorique, IRSAMC, 
  Universit\'e de Toulouse, CNRS, UPS, 31062 Toulouse, France}
\date{\today} 

\abstract{We study depletion dynamics of an open system of weakly interacting fermions with two-body random interactions. In this model  fermions are escaping from the high-energy one-particle orbitals,  that mimics the evaporation process used in laboratory experiments with neutral atoms to cool them to ultra-low temperatures.  It is shown that due to dynamical thermalization the system instantaneously adjusts to the new chemical potential and temperature,  so that occupation numbers of the one-particle orbitals always obey the Fermi-Dirac distribution.
In this way we are able to describe the evaporation process which leads to a significant cooling of particles remaining inside the system. We also briefly discuss the evaporation process in the SYK black hole model that corresponds to strongly interacting fermions.}


\maketitle

\section{Introduction}
The cooling of neutral atoms to micro-Kelvin  and further to nano-Kelvin temperatures is one of the most noticeable achievements of the modern physics that opened a new era of quantum technologies \cite{QT}. The main method used to cool atoms from micro to nano-Kelvin temperature is the evaporative cooling, where experimental setups are designed to remove the most hot atoms from an atomic  cloud. Although being quite successful the method has a drawback that one loses up to 99 percent of atoms to reach the temperature where Bose or Fermi atoms enter the degenerate state.

In this work we extend our previous studies \cite{105} of dynamical (or self-) thermalization in the Two-Body Random Interaction Model (TBRIM) 
introduced in Refs.~\cite{bohigas2,french2}. The model describes a system of $N$ spinless fermions distributed over $M$ one-particle orbitals, where fermions have two-body interactions with random matrix elements. We note in passing that at present the limiting case of this model with strongly interacting fermions got a renewed interest in the context of SYK black hole model for theory of quantum gravity \cite{kitaev2015,sachdev2015,polchinski,maldacena1}. 

It was shown in Ref.~\cite{105,frahm2018} that at moderate interaction strength TBRIM enters the regime of Quantum Chaos\footnote{This assumes the interaction strength to exceed some critical value which can be estimated by using  {\AA}berg's criteria \cite{aberg1}. Numerically the transition to Quantum Chaos \cite{haake,stockmann} is observed as a change of  the level-spacing statistics from the Poisson distribution to the Wigner-Dyson distribution when the interaction strength exceeds the above critical value \cite{epl25}.}
where its eigenstates  $|\Psi_E\rangle$ become self-thermalized, in spite of the fact that the system is isolated (i.e., not connected to any thermostat). This means, in particular, that occupation numbers $n_k=\langle \Psi_E |\hat{c}_k^\dagger\hat{c}_k | \Psi_E\rangle$  of the orbitals with energies $\epsilon_k$ obeys the Fermi-Dirac distribution,
\begin{equation}
\label{0}
n_k=\frac{1}{e^{\beta(\epsilon_k - \mu)} + 1}   \;,
\end{equation}
where the inverse temperature $\beta$ and the chemical potential $\mu$ are uniquely determined by the  eigenstate energy $E$ and the number of fermions $N$ in the system. Here we make the system open by introducing absorption of  particles from the most upper orbital. It is shown below that TBRIM preserves its self-thermalization property also in the presence of absorption.

Since we remove particles from  the high-energy orbital, the open TBRIM mimics the process of evaporative cooling of fermionic atoms. We argue in the work that the understanding of the mechanisms of self-thermalization allows to perform an optimization of the evaporative cooling. It involves, in particular, optimization of the rate  at which particles are absorbed in the system. We find that the optimal rate is actually determined by the self-thermalization rate in TBRIM. 
\begin{figure}[t]
\includegraphics[width=8.5cm,clip]{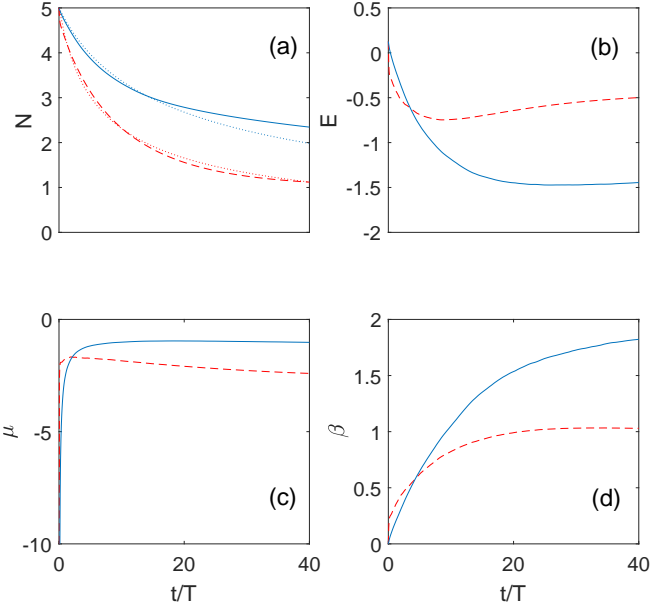}
\caption{The mean number of fermions $\bar{N}$ (a),  the mean energy $\bar{E}$ (b), the chemical potential $\mu$ (c), and the inverse temperature $\beta$  (d) as functions of the time for the depletion constant $\gamma=0.2$, blue solid lines, and $\gamma=2$, red dashed lines. Additional dotted lines in the panel (a) are Eq.~(\ref{66}). The system size is $M=12$ and initially we have $N_0=5$ fermions. Since we set $\hbar=1$ and $\overline{\epsilon_k^2}=1$ the characteristic time $T=2\pi$. }  
\label{fig0} 
\end{figure}   

\section{The model}
The Hamiltonian of the closed (isolated) TBRIM reads
\begin{eqnarray}
\label{1}
\widehat{H}^{(N)} =  \sum_{k=1}^M \epsilon_k \hat{c}_k^\dagger \hat{c}_k  
+\varepsilon \sum_{ijkl=1}^M J_{ij,kl}  \hat{c}_i^\dagger  \hat{c}_j^\dagger   \hat{c}_k \hat{c}_l  \;, 
\end{eqnarray}
where the orbital energies $\epsilon_k$ and the interaction (generally complex) constants $J_{ij,kl}$ are random numbers and we set the dispersion of all random entries to unity. The parameter $\varepsilon$ in the Hamiltonian (\ref{1}) controls the strength of two-body interactions and the super-index $N$ denotes the number of fermions in the system. The dimension of the Hilbert space is ${\cal N}_N=M!/N!(M-N)!$. In what follows we use $M=12$ and $\varepsilon=0.034$. We checked that for this value of $\varepsilon$ the TBRIM is chaotic and self-thermalized for $2\le N \le 5$, which are relevant for the numerical simulation reported below.

We describe the evaporation dynamics  by solving the master equation on the system density matrix ${\cal R}(t)$,
\begin{eqnarray}
\label{3}
\frac{d {\cal R}}{dt}=-i[\widehat{H},{\cal R}] +{\cal L}_{loss}({\cal R}) \;, 
\end{eqnarray}
\begin{eqnarray}
\label{4}
{\cal L}_{loss}({\cal R})=\frac{\gamma}{2} 
(\hat{c}^\dagger_M\hat{c}_M{\cal R}-2\hat{c}_M{\cal R}\hat{c}^\dagger_M 
+ {\cal R}\hat{c}^\dagger_M\hat{c}_M) \;, 
\end{eqnarray}
where $\gamma$ is the  absorption rate or depletion constant, i.e., the rate  at which particles are removed from the most upper one-particle orbital $\epsilon_M$. Notice that the density matrix ${\cal R}$ is defined in the extended Hilbert space given by direct sum of $N_0+1$ subspaces where $N_0=N(t=0)$ is the initial number of fermions. Correspondently, the Hamiltonian $\widehat{H}$ in Eq.~(\ref{3}) has  block structure with blocks given by Eq.~(\ref{1}). In what follows we consider $N_0=5$ where the total dimension of the Hilbert space is ${\cal N}=1+12+ 66+220+495+792=3003$. As the initial condition we choose an eigenstate of $\widehat{H}^{(N)}$ with $N=N_0$ from the middle of its spectrum. This choice corresponds to infinite effective temperature where occupation numbers of the natural orbitals are approximately equal. The quantities which we calculate are the occupation numbers
$n_k(t)={\rm Tr}[\hat{c}_k^\dagger\hat{c}_k {\cal R}(t)]$,
the mean number of fermions in the system,
$\bar{N}(t)=\sum_{k=1}^M n_k(t)$,
and the mean energy,
$\bar{E}(t)=\sum_{k=1}^M \epsilon_k n_k(t)$.
Typical behavior of $\bar{N}(t)$ and $\bar{E}(t)$ is exemplified in Fig.~\ref{fig0}(a,b) where the blue solid and red dashed lines refer to the case of a small $\gamma=0.2$ and a large $\gamma=2$, respectively.  In the subsequent two sections we analyze this behavior in some details.
\begin{figure}
\includegraphics[width=8.5cm,clip]{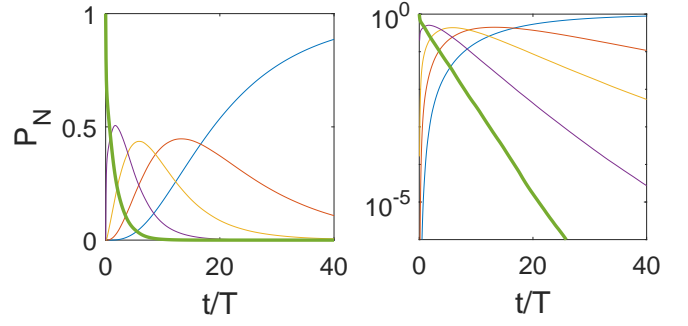}
\caption{Probabilities $P_N(t)$ to find $N$ particles at a given time $t$ in the linear, left panel, and logarithmic, right panel, scales. Different curves refer (from top to bottom at $t/T=40$) to $N=1,2,3,4,5$, respectively. The survival probability (i.e., the probability to find initial number of fermions) is marked by the thick line. The value of the depletion constant $\gamma=2$.} 
\label{fig1}
\end{figure}

\section{Depletion dynamics}
We begin with the depletion dynamics. Similar to the total Hamiltonian the density matrix ${\cal R}(t)$ has the block structure where each block is associated with the fixed number of fermions in the system.  This gives another expression for the mean number of  particles, 
\begin{eqnarray}
\label{5}
\bar{N}(t) =\sum_{N=1}^{N_0} N P_N(t) \;, \quad P_N(t)={\rm Tr}[{\cal R}^{(N)}(t)]/N \;,
\end{eqnarray}
where $P_N(t)$ are interpreted as probabilities to find $N$ particles in the system at a given time $t$. Probabilities $P_N(t)$ are shown in Fig.~\ref{fig1} in the linear and logarithmic scales for $\gamma=2$. It is seen that $P_N(t)$ with $N>1$ show asymptotic exponential decay while $P_1(t)$ approaches unity. This is consistent with the expectation that the steady-state solution of the master equation (\ref{3}) corresponds to a single fermion. It is also seen in Fig.~\ref{fig1} that relaxation to this steady-state is a cascade-like process where `children' cascade takes essentially longer time than  `parent'  cascade. In the other words, depletion rate decreases proportionally to the number of particles left in the system. In the next paragraph we give a formal explanation for this intuitively expected result and quantify the dependence $\bar{N}(t)$.

One gets a useful insight in the depletion process by analyzing the survival probability, i.e., probability to find the initial number of particles. The survival probability has no parent cascade and can be calculate in a simpler way, namely, by solving the Schr\"odinger equation with the non-Hermintian Hamiltonian $\widehat{H}_{eff}$, 
\begin{eqnarray}
\label{6}
\widehat{H}^{(N)}_{eff}
= \sum_{k=1}^M \left(\epsilon_k -i\frac{\gamma}{2} \delta_{k,M}\right) \hat{c}_k^\dagger \hat{c}_k  
+\varepsilon \sum_{ijkl=1}^M J_{ij,kl}  \hat{c}_i^\dagger  \hat{c}_j^\dagger   \hat{c}_k \hat{c}_l  \;, 
\end{eqnarray} 
which is obtained  from the Herminia Hamiltonian (\ref{1}) by prescribing imaginary energy to the $M$th orbital. Then the norm of the wave function exactly corresponds to probability to find the initial number of particles in the system.  The introduced non-Herminia Hamiltonian is related to the problem of quantum chaotic scattering  \cite{Soko88,lehmann,sommers,savin,50}. Considering the Hamiltonian matrix in the Fock basis we find the number of complex diagonal elements to be given by $Q=N{\cal N}_N/M$, which can be interpreted as the number of decay channels. It is known that short-time dynamics of the survival probability in chaotic scattering is the exponential decay, $P_N(t)=\exp(-\nu t)$,  with increment $\nu$ proportional to ratio of  the number of channels to the matrix size \cite{savin,50}. Adopting this result to our problem with cascade dynamics (where the number of channels and the matrix size are changing from parent to children cascade) we obtain
\begin{eqnarray}
\label{66}
\bar{N}(t)=N_0\exp\left[-\alpha \bar{N}(t)\; t\right] \;,
\end{eqnarray}
where the coefficient $\alpha$ is a function of  the depletion constant $\gamma$.  The solution of the nonlinear equation (\ref{66}) is depicted in Fig.~\ref{fig0}(a) by dotted lines. It is in a reasonable agreement with the straightforward simulation of the depletion dynamics.
\begin{figure}
\includegraphics[width=8.5cm,clip]{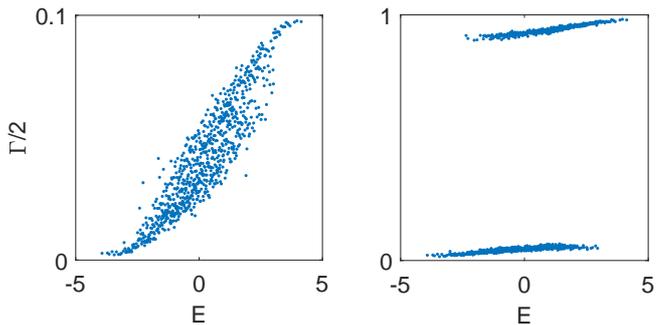}
\caption{Eigenvalues of the Hamiltonian (\ref{6}) in the complex plane for $\gamma=0.2$, left panel, and $\gamma=2$, right panel.} 
\label{fig2}
\end{figure}

At the next step we discuss the dependence of the increment  $\nu$ for the exponential decay of the survival probability on the depletion constant $\gamma$. One may naively expect that $\nu$ [and, hence, the coefficient $\alpha$ in Eq.~(\ref{66})] linearly depends on $\gamma$. However, this is true only if the depletion constant is smaller than the rate of self-thermalization, while in the opposite case $\nu$ actually decreases  with an increase of $\gamma$. This effect, which is often referred to as the Zeno effect \cite{zeno}, has a simple explanation in terms of the energy spectrum of the Hamiltonian (\ref{6}),
\begin{eqnarray}
\label{8}
\widehat{H}_{eff} |\phi_j\rangle={\cal E}_j |\phi_j\rangle  \;, \quad {\cal E}_j =E_j-i\frac{\Gamma_j}{2} \;.
\end{eqnarray}
Indeed, for small $\gamma$ the complex energies ${\cal E}_j$ are located near the real axis, see Fig.~\ref{fig2}(a), and the mean value $\bar{\Gamma}$ of the resonance widths  is proportional to $\gamma$, $\bar{\Gamma}\sim \gamma$.  As $\gamma$ is increased, the cloud of eigenenergies splits into two groups\footnote{In the context of quantum chaotic scattering this effect is discussed in Refs.~\cite{lehmann,sommers}.},
see Fig.~\ref{fig2}(b). For the first group $\bar{\Gamma}$ remains to be proportional to $\gamma$ while for the the second group we have $\bar{\Gamma}\sim 1/\gamma$ , as it is easy to show by using the first-order perturbation theory. Thus, with an increase of the depletion constant $\gamma$ the increment $\nu=\bar{\Gamma}$  shows a maximum at some critical value $\gamma_{cr}$, which for the parameters of Fig.~\ref{fig1} corresponds to $\gamma_{cr}\approx 1.2$. We mention that this critical value of the depletion constant can be used as  unambiguous definition of the self-thermalization rate in the considered system of weakly-interacting fermions.

\section{Thermalization dynamics}
In this section we provide numerically evidence that TBRIM remains to be self-thermalized also in the presence of particle loss. The numerical confirmation of this statement is given below.
\begin{figure}
\includegraphics[width=8.5cm,clip]{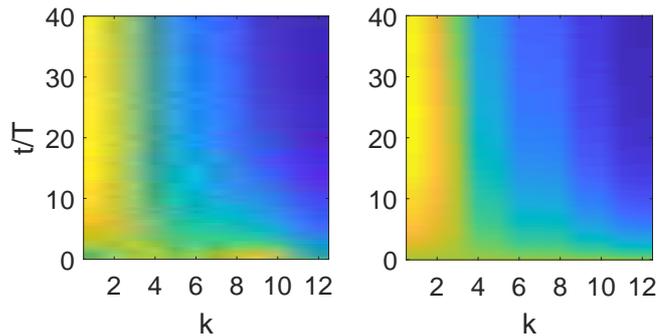}
\caption{Occupation numbers of the orbitals, left panel, as compared to the Fermi-Dirac distribution with $\mu=\mu(t)$ and $\beta=\beta(t)$ taken from  Fig.~\ref{0}(c,d), right panel.} 
\label{fig5}
\end{figure}   

\begin{figure}
\includegraphics[width=8.0cm,clip]{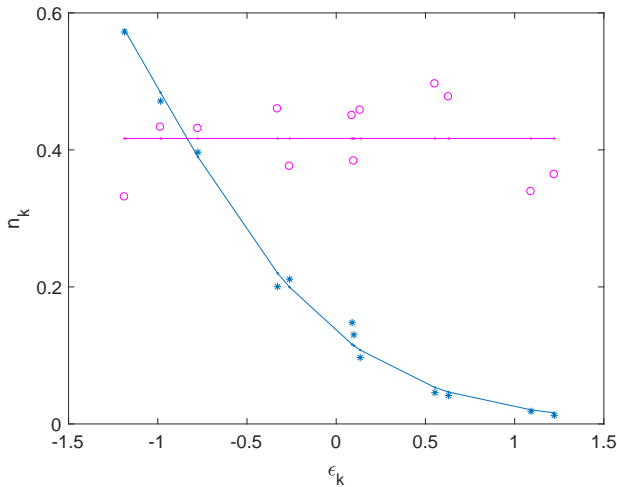}
\caption{Occupation numbers of the orbitals at the beginning (open circles) and the end (asterisks) of numerical simulation as the function of orbital energies. Solid lines are the Fermi-Dirac distribution for $\beta=0$ and $\beta\approx 1.8$ that correspond to the initial and final temperature of the system.} 
\label{fig6}
\end{figure}

First, using  numerical data for the mean number of particle $\bar{N}$ and the mean energy $\bar{E}$ depicted in Fig.~\ref{fig0}(a,b) we solve the system of two nonlinear algebraic equation on the chemical potential $\mu$ and the inverse temperature $\beta$,
\begin{equation}
\label{9}
\bar{N} = \sum_{k=1}^M\frac{1}{e^{\beta(\epsilon_k - \mu)} + 1}  \;,\quad
\bar{E} = \sum_{k=1}^M\frac{\epsilon_k}{e^{\beta(\epsilon_k - \mu)} + 1}  \;.
\end{equation}
The solution is shown in the panels (c) and (d)  where, as before, the solid and dashed lines refer to $\gamma=0.2$ and $\gamma=2$, respectively. Second, using the obtained $\mu$ and $\beta$ we calculate the occupation numbers according to Eq.~(\ref{0}) and compare them with actual occupation numbers  calculated  as $n_k(t)={\rm Tr}[\hat{c}_k^\dagger\hat{c}_k {\cal R}(t)]$, see Fig.~\ref{fig5}. (From now on we focus on the case $\gamma=0.2$.) The observed agreement confirms that we indeed have the true thermalization where the notion of temperature is absolutely meaningful, in spite of the absence of any thermal bath. From Fig.1(d) we clearly see that during evaporation the temperature of the system decreases with time,

In addition to Fig.~\ref{fig5}, Fig.~\ref{fig6} shows occupation numbers at the beginning and the end of numerical simulation as the function of orbital energies. Remarkably,  occupations of two lowest-energy orbitals get increased. Thus, for a larger system size one may expect that $n_k$ of a few  lowest-energy orbitals approach unity, i.e., we enter the degenerate state.

\section{Case of SYK black hole model}

Due to a significant recent interest to the SYK model \cite{kitaev2015,sachdev2015,polchinski,maldacena1} we briefly discuss the evaporation process in this model which corresponds to vanishing orbital energies  $\epsilon_k=0$  in the Hamiltonian (\ref{1}). Clearly, in this case of strongly interacting fermions we cannot appeal to the Fermi-Dirac distribution (which is strictly derived for non-interacting fermions) and, hence, the notion of temperature is not defined. However, we still can address the depletion dynamics.

\begin{figure}
\includegraphics[width=8.0cm,clip]{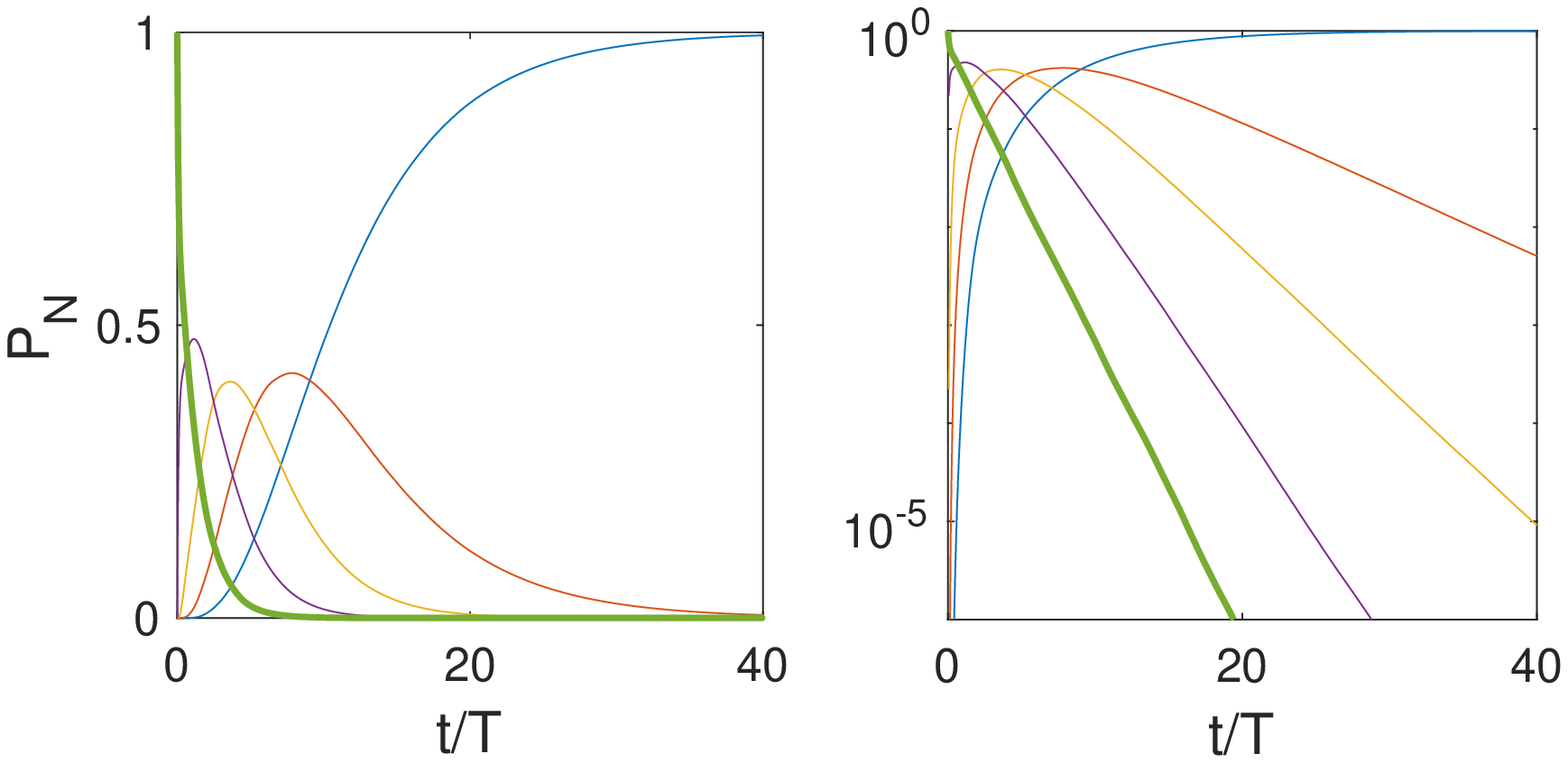}
\includegraphics[width=8.0cm,clip]{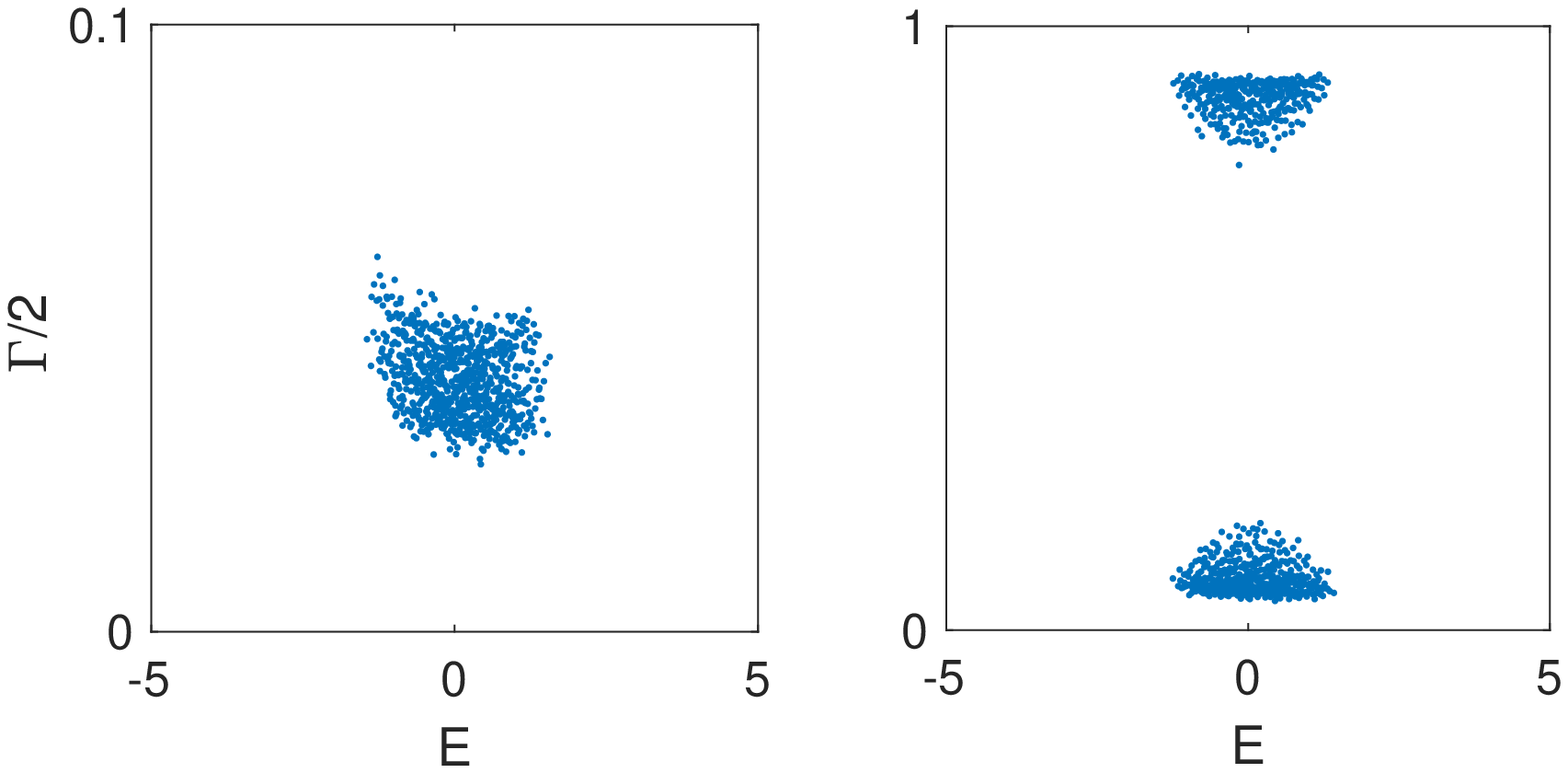}
\caption{Top panels: the same as in Fig.~\ref{fig1} yet for the SYK case where one-particle energies $\epsilon_k$ in the Hamiltonian (\ref{1}) are set to zero. Bottom panels: the same as Fig.~\ref{fig2} yet for the SYK case where $\epsilon_k$ in the Hamiltonian (\ref{6}) are set to zero. Parameter $\varepsilon$ in both Hamiltonians is $\varepsilon=0.034$} 
\label{fig7}
\end{figure}

As for the previous TBRIM case we assume the absorption to take place only from one orbital. Upper panels in Fig.~\ref{fig7} shows probabilities $P_N(t)$ for  $\gamma=2$ in the linear and logarithmic scales. Similar to the TBRIM case there is an exponential decay of probability to find the number of fermions $N>1$ while the probability $N=1$ approaches unity. The decay rate of the survival probability is seen to be slightly larger than in the TBRIM case,  which is consistent with distribution of eigenvalues of the non-Hermitian Hamiltonian -- compare Fig.~\ref{fig2}(b) and Fig.~\ref{fig7}(d).  For the sake of completeness we also display in Fig.~\ref{fig7}(c) the eigenvalue distribution for $\gamma=0.2$.  Similar to the TBRIM case this compact cloud of eigenvalues is found to separate into two clouds at $\gamma \approx 1.2$. Thus for both TBRIM and SYK models the depletion rate is maximized  at the same value of $\gamma_{cr}$.  

\section{Conclusions}

We analyzed the process of evaporative cooling in the system of weakly interacting fermions. This process has a competition between depletion, where particles are removed from high-energy one-particle orbitals (generalization of the results onto the case of more than one decaying orbital is straightforward) and self-thermalization, which repopulates these orbitals.  We especially stress the importance of the latter process and it is generally a hard problem to find conditions for emergence of self-thermalization in a given system of interacting particles. For the isolated TBRM these conditions have been analyzed in our previous  works  \cite{105,frahm2018}. In the present work we showed that the self-thermalization in TBRIM takes place also in the presence of particle loss (absorption). Thus the TBRIM becomes a paradigmatic model for investigation of different aspects of self-thermalization in closed and open systems.

We studied the depletion dynamics  in the open TBRM by mapping the problem to quantum chaotic scattering. It was shown, in particular, that the number of lost particles  is not a monotonic function of the depletion constant $\gamma$ but has a pronounced maximum at some $\gamma_{cr}$. Since the depletion constant can be varied in laboratory experiments, this result suggests  a method for measuring the self-thermalization rate by finding the critical $\gamma_{cr}$ where the increment $\nu$ for exponential decay of the survival probability is maximized.

Since we remove particles from the high-energy orbitals, the depletion process results in a decrease of the system energy and, as a consequence,  in the temperature drop for the remaining particles. We stress that efficiency of this cooling mechanism is not just proportional to particle loss, as one might naively expect, but is an involved function of the depletion constant $\gamma$ (which can be varied in a laboratory experiment),  the self-thermalization rate $\gamma_{cr}$ (which is an internal property of the system), and the duration of the evaporation process (which can be limited by some reason). An example is given   in Fig.~\ref{fig0}(a,d). It is seen that within the same time interval we reached lower temperature for $\gamma=0.2$ than for $\gamma=2$, in spite of the fact that in the former case we lost less particles than in the later case.    
  

\section{Acknowledgements}
For DLS this work was supported in part by the Programme Investissements  
d'Avenir  ANR-11-IDEX-0002-02,  reference  ANR-10-LABX-0037-NEXT  (project  THETRACOM).



\end{document}